\documentclass[12pt]{article}
\usepackage{epsfig}
\usepackage{amsmath}
\usepackage{hhline}
\usepackage{amssymb}
\usepackage{times}

\newlength{\dinwidth}
\newlength{\dinmargin}
\def\Journal#1#2#3#4{{#1} {\bf #2}, #3 (#4)}

\def\NIMA{{\em Nucl. Instrum. Methods} A}
\def\NPB{{\em Nucl. Phys.} B}
\def\PLB{{\em Phys. Lett.}  B}

\def\PRD{{\em Phys. Rev.} D}
\def\ZPC{{\em Z. Phys.} C}
\def\EPJC{{\em E. Phys. J.} C}
\def\be{\begin{equation}}
\def\ee{\end{equation}}
\def\bea{\begin{eqnarray}}
\def\eea{\end{eqnarray}}
\def\etal{{\it et~al.}}
  \catcode`\"=\active \def"{\"}
  \def\lca{\,\raise.5ex\hbox{\rlap{$<$}} \lower.6ex\hbox{$\sim$}\,}
  \def\gca{\,\raise.5ex\hbox{\rlap{$>$}} \lower.6ex\hbox{$\sim$}\,}
  \def\macenter{\hbox{\raise.8ex\hbox{$-$}\kern-7pt \hbox{$\wedge$}}}
  \def\sqr#1#2{{\vcenter{\vbox{\hrule height.#2pt
          \hbox{\vrule width.#2pt height#1pt \kern#1pt\vrule width.#2pt}
          \hrule height.#2pt}}}}
  \def\grad{^{\circ}}     % degree
  \def\nb{\,\rm nb}
  
  \def\pt{p_\perp}
  \def\ptrel{p_{T, rel}^\mu}

\begin{document}  
% The rest
\newcommand{\gev}{\,\mbox{GeV}}
\newcommand{\result}{0.176~\pm~0.016~(stat.)~\,^{+0.026}_{-0.017}~(syst.)\, \nb}
\newcommand{\totep}{7.1 \pm 0.6  \,^{+1.0}_{-0.7}\pm 1.1 \, \nb}
\newcommand{\totgp}{111\pm 10  \,^{+16}_{-11}\pm 17 \, \nb}
\newcommand{\rtotep}{7.1 \pm 0.6  (stat.) \,^{+1.5}_{-1.3} (syst.) \, \nb}
\newcommand{\rtotgp}{111~\pm~10~(stat.)~\,^{+23}_{-20}~(syst.)~\,~\nb}
\newcommand{\theory}{0.104~\pm~0.017\,~\nb}
\newcommand{\xsbb}{$\sigma(ep~\!\rightarrow~\!b~\!\bar{b}~\!X~\rightarrow~\!\mu~\!X')\ $}
\begin{titlepage}

\noindent
DESY 99-126\\

\vspace{2cm}

\begin{center}
\begin{Large}

{\bf  Measurement of Open  Beauty Production at HERA}

\vspace{2cm}

H1 Collaboration

\end{Large}
\end{center}

\vspace{2cm}

\begin{abstract}
\noindent  The first observation of open $b$ production in
$ep$ collisions is reported.  An event sample containing muons and
jets has been selected which is enriched in semileptonic $b$ quark
decays.  The visible cross section \xsbb for $Q^2~<~1~\gev^2$, $0.1 <
y < 0.8$ is measured to be $\result $ for the muons to be detected in
the range $35\grad < \theta^\mu < 130\grad$ and $\pt^\mu > 2.0 \gev$
in the laboratory frame.  The expected visible cross section based on
a NLO QCD calculation is $\theory$. The cross sections for
electroproduction with $Q^2<1~\gev^2$ and photoproduction are derived
from the data and found to be $\sigma(ep\rightarrow e
b\bar{b}X)=\rtotep$ and $\sigma(\gamma p\rightarrow b\bar{b} X)=
\rtotgp$ at an average $\langle W_{\gamma p}\rangle \sim 180 \gev$,
respectively.
\end{abstract}

\vspace*{30mm}
\begin{center}
  {\it  Submitted to Physics Letters B}
\end{center}

\end{titlepage}
\noindent
%\small
%   H1AUTS  Author list by names, no. of authors  347
%           status: 23/06/99   09.16.06
 C.~Adloff$^{33}$,                %WUPP-ST                  Adloff             
 V.~Andreev$^{24}$,               %LPI -PD                  Andreev            
 B.~Andrieu$^{27}$,               %ECPL-PD                  Andrieu            
 V.~Arkadov$^{34}$,               %ZEUT-PD    10/96         Arkadov            
 A.~Astvatsatourov$^{34}$,        %ZEUT-ST     2/98         Astvatsatourov     
 I.~Ayyaz$^{28}$,                 %PARI-ST       5/96       Ayyaz              
 A.~Babaev$^{23}$,                %ITEP-PD                  Babaev             
 J.~B\"ahr$^{34}$,                %ZEUT-PD                  Baehr              
 P.~Baranov$^{24}$,               %LPI -PD                  Baranovp           
 E.~Barrelet$^{28}$
% ...

%\footnotesize
\noindent
%     H1 Institutes as appearing on publications
% ...
$ ^{23}$ Institute for Theoretical and Experimental Physics,
          Moscow, Russia \\
 $ ^{24}$ Lebedev Physical Institute, Moscow, Russia$^{f,k}$ \\
% ...
$ ^{27}$ LPNHE, \'{E}cole Polytechnique, IN2P3-CNRS, Palaiseau, France \\
 $ ^{28}$ LPNHE, Universit\'{e}s Paris VI and VII, IN2P3-CNRS,
          Paris, France \\
% ...
$ ^{33}$ Fachbereich Physik, Bergische Universit\"at Gesamthochschule
          Wuppertal, Wuppertal, Germany$^a$ \\
 $ ^{34}$ DESY, Zeuthen, Germany$^a$ \\
% ...
$ ^a$ Supported by the Bundesministerium f\"ur Bildung, Wissenschaft,
        Forschung und Technologie, FRG,
        under contract numbers 7AC17P, 7AC47P, 7DO55P, 7HH17I, 7HH27P,
        7HD17P, 7HD27P, 7KI17I, 6MP17I and 7WT87P \\
 $ ^b$ Supported by the UK Particle Physics and Astronomy Research
       Council, and formerly by the UK Science and Engineering Research
       Council \\
% ...
$ ^f$ Supported by the Deutsche Forschungsgemeinschaft \\
% ...
$ ^k$ Supported by Russian Foundation for Basic Research 
       grant no. 96-02-00019 \\
% ...
% $ ^{m}$ Foundation for Polish Science fellow \\

\newpage

\normalsize
\subsection*{Introduction}
\noindent 
The study of  heavy quark production in electron--proton scattering 
provides an important testing ground for QCD.
  Measurements of open charm production 
\cite{heracharm} at the electron-proton collider, HERA,  have been 
shown to be reasonably well described  by next to leading order (NLO) QCD 
calculations \cite{ellis,frixione,smith,harris} based on the photon-gluon 
fusion mechanism.  It is essential to extend these observations and tests of 
QCD to the production of heavier quarks.

In the past, the leptoproduction of open $b$ quarks has been too 
difficult to observe due to the small production cross section. 
At fixed target experiments only upper limits on the leptoproduction 
cross section have been published \cite{EMC,BCDMS,BFP}   
\footnote{EMC also derived a cross section based on the 
observation of one clear event \cite{EMC}.}.
In this paper, the first  observation of  $b$ quark 
leptoproduction is presented in the range $Q^2$ less than 1 $\gev^2$, where 
$Q^2$ is the four-momentum transfered from the electron to 
the proton. This corresponds to the exchange of an almost 
real photon (photoproduction) i.e. $Q^2 \sim 0$.   
 
The data were collected  at the HERA collider with the H1 detector
during 1996 and correspond to an integrated luminosity of
${\cal L}=6.6\,{\rm  pb^{-1}}$. 
 The  production of  $b$ quarks is observed through their semileptonic 
decay to muons.  The production cross section is measured
in a kinematic  range determined by the properties of the muon from 
the heavy quark decay and is
compared with the prediction of a NLO QCD calculation 
for partons \cite{frixione} to which  a model
 for the hadronisation and semileptonic decays is added.

\subsection*{Detector Description}
 
Only a short description of the detector components which play a major
role in this analysis is given here; a detailed description of the
detector and its trigger capabilities can be found elsewhere
\cite{detector}.  Charged particles are measured by two cylindrical
jet drift chambers (CJC) \cite{cjc}, mounted concentrically around the
beam line inside a homogeneous magnetic field of 1.15 Tesla, yielding
measurements of particle momenta in the polar angular
range\footnote{H1 uses a right-handed coordinate system with the
  $z$-axis pointing in the direction of the proton beam (forward) and the
  $x$-axis pointing towards the centre of the storage ring.}  of
20$^{\circ}$ to 160$^{\circ}$.  Two double layers of cylindrical
multiwire proportional chambers \cite{mwpc} with pad readout for
triggering purposes are positioned inside and in between the two drift
chambers.  The tracking detector is surrounded by a fine grained
liquid argon calorimeter \cite{calo}, covering a polar angle range of
4$^{\circ}$ to 154$^{\circ}$, consisting of an electromagnetic section
with lead absorbers and a hadronic section with steel absorbers.
Muons are measured as tracks in the CJC and identified by demanding a
good link to a track in the instrumented iron return yoke which
surrounds the superconducting coil.  For events from deep--inelastic
scattering (DIS) at higher $Q^2$ the scattered electron is detected by
a ``spaghetti--type'' lead-scintillator-fibre calorimeter (Spacal)
\cite{spacal} which covers the backward region of the detector up to a
polar angle of $177.8\grad$.

\subsection*{ Analysis Procedure}
The production of $b$ quarks is detected by looking for muons from the 
semileptonic  decay of a $b$ hadron inside jets from the hadronisation 
of the $b$ quark.
A sample of events enriched in semileptonic decays of $b$ quarks is selected 
by demanding that each event has at least 2 jets
and that at least one of the jets should contain a muon.  The 
jets are built by using both the calorimetric and track information 
and are identified by using a cone algorithm \cite{cone} with a cone 
radius $r=\sqrt{(\Delta\eta)^2+(\Delta\phi)^2}<1$, 
where $\phi$ is the azimuthal angle and $\eta$ is the pseudorapidity 
and  with transverse energy to the beam direction, $E_T$, of more than $6 \gev$.
The muon is required to be  
in the central region of the detector ($35\grad <
\theta^\mu < 130\grad$) and to have a transverse momentum with respect
to the beam  direction,  $\pt^\mu$, of more than $2\gev$.
The selected events have a mean track multiplicity of 16.5 and a minimum 
of at least 5 tracks per event is required.
 The events are triggered by requiring 
the existence of a muon candidate  in coincidence 
with tracks. The trigger efficiency to select 
these events is determined directly from the data to be
$85.5 \pm 1.6 \%$ and is found to be constant for  $\pt^\mu >2\gev$. 
To suppress contributions from DIS at higher 
$Q^2$ the events which contained an electron candidate with 
more than $4\ \gev$ energy deposited 
 in the Spacal are rejected. To avoid regions of high radiative corrections 
and to eliminate the remaining DIS events, the  value of $y$ was 
limited to the    range $0.1<y<0.8$. Here $y$ is 
$q\cdot~\!P/\ell\cdot~\!P$ where $q$, $P$ and $\ell$ are the
 four momenta  of the virtual photon, incident proton and lepton, respectively.
 It is calculated using the Jacquet-Blondel method 
\cite{jacquet}.
These conditions limit the data sample to photoproduction 
events with $Q^2<1\gev^2$;  the mean $Q^2$ is $6\cdot 10^{-2}\gev^2$. 
In total 927 events  fulfil these selection criteria.

 Beauty is  separated from charm and other backgrounds 
on a statistical  basis  using the  transverse  momentum of the   muon,
$\ptrel$, measured relative to the thrust axis of the  jet which contains it.
  The thrust axis is defined as the direction such that 
$$ T = \max \frac{\sum |p_i^L|}{ \sum|p_i|}, $$
where the sums run over all the neutral and charged  particles in the jet 
with the exception of the muon. Here, $p_i$ is the
momentum and  $p_i^L$ its longitudinal component with respect to the
thrust axis.
Since the $b$ quark is heavy,  it is expected that
the $\ptrel$ spectrum is harder for  events originating 
from $b$ quark decays than for decays from lighter quarks.
For the determination of the beauty cross section, 
the $\ptrel$  spectrum is used  for a combined fit of the
different contributions to the selected data sample. The shapes
of the $b$ and $c$ contributions to the $\ptrel$  are taken from the
 {\sc Aroma}  Monte Carlo \cite{aroma22} event
generator, which is based on leading order (LO) QCD matrix elements,
with the photon directly interacting with the partons in the proton
and additional initial and final state parton showers.

\begin{figure}[htb]
  \begin{center}
    \mbox{\epsfig{file=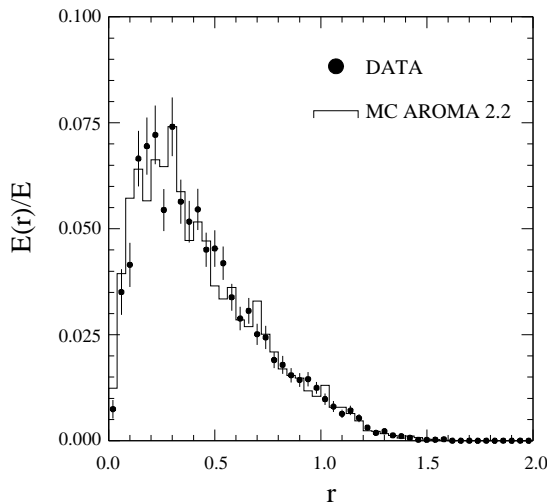,width=0.45\linewidth,clip=}} 
    \caption{ \small Fractional energy flow within jets as a function of 
the distance to the jet axis. This fraction is the ratio of the energy in an annulus at radius
$r$ to the total energy in the jet. The data points (filled dots) correspond to a data sample
 of mainly  charm decays and the full histogram to a Monte 
Carlo simulation where only charm quarks are generated. Both the data and 
Monte Carlo are normalized so that the integral of the distributions equals
 one. }
    \label{fig:eflow}
  \end{center}
\end{figure}
The method relies on a good understanding of the  energy flow
within  jets, for jets originating from heavy quarks,
 and on a correct determination of the muon 
misidentification probability.
To check that the Monte Carlo simulation  describes the energy
distribution within  jets, a data sample enriched in charm quarks
is studied, where events with at least one jet and at least one
$D^*$ candidate  are selected.
The $D^{*\pm}$ candidate is identified through the
decay $ D^{*+}\rightarrow {D^0} \pi^+\rightarrow
K^- \pi^+ \pi^+ $ and its charge conjugate.
  This gives a sample of events with  20\% background which has been 
verified to have a similar distribution to charm.
In each of these events jets are identified using the same 
selection criteria as in the muon sample.
The energy flow within the jet has been compared  to 
the energy flow as given by the {\sc Aroma} Monte Carlo event
generator for charm events with a $D^*$.  There is a good agreement
between the data and the prediction from the Monte Carlo 
(see figure \ref{fig:eflow}).
 From this and many other such comparisons 
 it is concluded that the Monte Carlo simulation
 reliably models the data distributions.

\begin{figure}[htb]
  \begin{center}%\vspace*{-4mm}
          \mbox{\epsfig{file=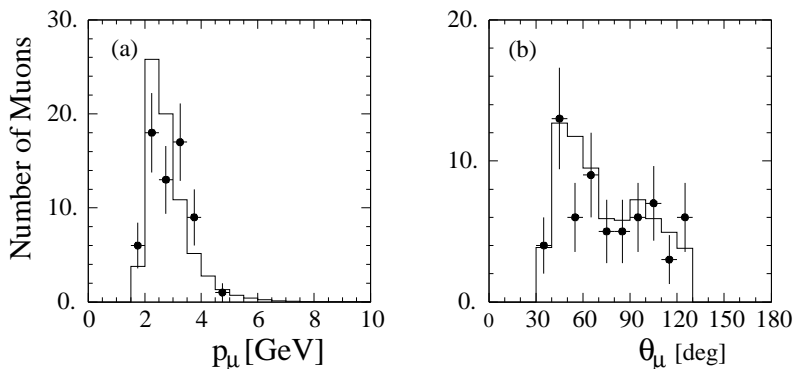,width=0.65\linewidth,clip=
}}\vspace*{-4mm}
\caption{ \small Momentum  (a) and polar angle
(b)  distributions 
for fake muons in  a pion sample from the   
decay  $K^0_S\rightarrow\pi^+\pi^-$.  The  filled points show the fake
muon yield  as measured   in the data.   The  solid  histogram  gives  the
estimate of the muon yield as obtained by assigning ${\cal  P}^{\mu}_h(p,\theta)$ as a weight
to every pion  and  summing the weights  over the  entire pion sample. The
measured yield   is  64 fake  muons; the   estimate amounts  to 69.8.}
    \label{fig:verifypions}
  \end{center}
\end{figure}

The selected sample contains muons originating from the semileptonic 
decays of $b$ and $c$ quarks as well as from light
 hadrons which are misidentified as muons, due to decay or hadronic energy leakage (punchthrough).
To obtain the background contribution 
 originating from misidentifying 
 a light hadron as a muon,
it is essential to know 
the probability ${\cal   P}^\mu_h(p,\theta)$ for this to occur.
Large samples of single 
pion, kaon and proton tracks are passed  through the full detector 
simulation and the fraction measured as muons is determined.
  From these, the  misidentification probability functions ${\cal
  P}^\mu_h(p,\theta)$, where $h=\pi, K ,
p$, are parameterised as a function of the momentum and
the polar angle of the hadron.
These functions  vary with the polar angle but  do not
exceed $6\cdot 10^{-3} $ in the case of pions and $2\cdot 10^{-2}$ in
the case of kaons. For protons it  is
found to be below $2\cdot 10^{-3}$.
 These probability functions are verified in the data  by
studying $K^0_S$ and $\phi$ decays,  as a source of real  pions and kaons,  
respectively.
The distributions of fake muons  agree both in shape and absolute 
magnitude (figure \ref{fig:verifypions})
 with those predicted by applying  the probability functions to 
the pions  from $K^0_S$ decays.
For kaons from $\phi$ decays, 8 fake muons are observed compared to 7.8 
predicted. These
observations  show that the probability  functions are well determined.

\par The knowledge of the ${\cal  P}^{\mu}_h(p,\theta)$ functions allows
 the background from hadrons, which are falsely  identified as muons
in the $\ptrel$ spectrum, to be calculated
   in shape and absolute magnitude from the data. For this
purpose, events with at least two
jets  with $E_T~>~6~\gev$ each,
 but without the muon requirement, are selected. 
The expected background in the $\ptrel$  distribution is calculated
from all hadrons which pass the $\pt^\mu$ and polar angle requirements.
The assignment of a hadron as a pion, kaon or proton is made using
the {\sc Jetset}\cite{jetset} specification of the fractions of these 
particles which is in agreement with  measurements by the SLD and DELPHI 
collaborations
\cite{fractions}.  The {\sc Jetset} pion and kaon fractions are 
varied by $\pm 0.05$, covering the largest deviation of {\sc Jetset} from these
measurements.
This variation  corresponds to about  $15\%$ relative variation of the 
kaon fraction. The result of this variation is included in the systematic 
error for this  background determination.

\subsection*{Results}
\begin{figure}[h]
  \begin{center}%\vspace*{-4mm}
          \mbox{\epsfig{file=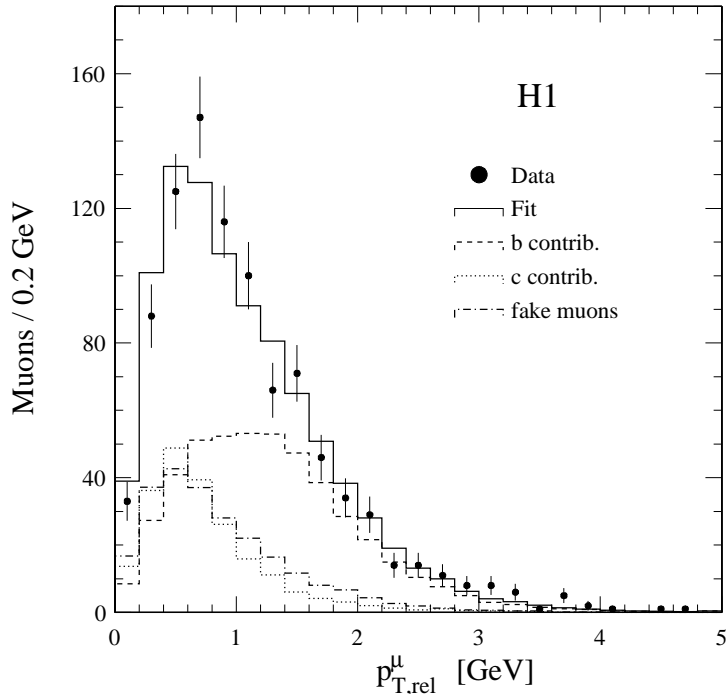,width=0.6\linewidth,clip=}}
\caption{ \small The measured $\ptrel$ distribution in the data 
and the  fitted  sum (solid line) of the contributions of 
beauty (dashed line), charm (dotted line) and the fixed fake muon background
(dashed-dotted line).}\label{fig:fitptrel}
  \end{center}
\end{figure}

The observed $\ptrel$ distribution is shown in figure \ref{fig:fitptrel}
 together 
with the $b$ signal and the different background contributions.
  The signal and backgrounds 
are obtained from the combined fit of the $b$ and $c$ contributions, using 
the shapes from the {\sc Aroma} Monte Carlo, together with the fake muon
 background determined as described above.
 The relative composition   of  the data  sample amounts  to $f_b =50.8
\pm 4.9 \%$  (beauty), $f_c = 22.4 \pm  5.0 \%$ (charm), and $f_{fake}=
25.9  \%$   (background, fixed). 
The uncertainty of the latter is included in the systematic error which is deduced by changing the 
pion and kaon fractions as described above. 

The visible electroproduction cross section for $b$ quarks, 
 determined from the number of muon
events    attributed to $b$ quark decays, is measured to be:
\bea
\sigma_{vis}(ep\rightarrow b\overline{b}X\rightarrow \mu X')
 &= \result \nonumber \eea
 in the kinematic range $Q^2 <  1\gev^2$, $0.1 < y <  0.8$,
$\pt^\mu~>~2.0~\gev$ and 
$35\grad < \theta^\mu < 130\grad$,
  where the first uncertainty is statistical and the second
systematic.

The acceptance is determined from the {\sc Aroma} Monte Carlo simulation
to be $24\%$, of which the efficiency for muon identification is 56\%
 and the probability to 
reconstruct two jets each with $E_T>6~\gev$ is 42.9\%. The full analysis has 
been repeated using   the {\sc
  Herwig   5.9}~\cite{herwig} and {\sc Rapgap}~\cite{rapgap}
  Monte    Carlo  simulations assuming both direct and resolved 
production of the $c$ and $b$ quarks. In addition,  different $E_T$ cuts for the jets
 and different values of the Peterson fragmentation parameter, 
$\varepsilon$, for both  the $c$ and $b$ fragmentation have been used.
The variation of the cross section 
with the different simulations amounts to  $\pm 7.1\%$ and this is taken to be the
systematic error due to the uncertainties in the Monte Carlo.
The systematic uncertainty  due to the muon reconstruction is $+6\%$. 
The systematic error due to the uncertainty  ($\pm 4\%$) in the
hadronic  energy scale of the calorimeter is 
 $\pm 4.9\%$. 
The systematic uncertainty from the luminosity calculation is $\pm 1.8\%$.
The error due to the background shape and magnitude is found to be 
$^{+9.5}_{-3.6} \%$ by changing  the
assumption on the $\pi, K, p$ composition in the measured hadron
sample, as described above, as well as by using  event samples 
selected by different triggers of the experiment to estimate the background.

To estimate the  theoretically expected cross section, the  
 NLO QCD calculation by Frixione {\it et al.} (FMNR) \cite{frixione}
is used, with 
$m_b=4.75 \gev$ and the MRSG \cite{mrsg} and GRV-HO \cite{grv}
 structure functions for the proton and photon, respectively.
This program supplies the kinematics of the $b$ quarks. In order to  
determine the expectation for the kinematic range where the  measurement 
is performed, the $b$ quarks are fragmented to mesons and then allowed to decay
semileptonically. For the $b$ quark fragmentation the Peterson  
parameterisation \cite{peterson} is used with 
$\varepsilon_b = 0.006\pm 0.003$ \cite{epsilonb}.
If the fragmentation is done by simply scaling the $b$ quark four vector by the
 factor generated according to the 
 Peterson function the expected cross section in the
visible range is \xsbb$=0.118 \, \nb$.
However, if the hadronisation of the $b$ quark 
is done together  with the generation of  a light quark pair
 then the expected cross section in the
visible range is \xsbb$=0.089 \, \nb$.
This difference is due to a softer  transverse momentum 
 distribution of the $b$ hadron in the 
latter case, which influences the $ \pt^\mu$ spectrum where a cut is applied.
For the theoretical  expectation the average of the above values is taken 
and the difference is taken to be  the systematic uncertainty.  
A variation  in the renormalisation or factorisation scale, which is defined
as $\mu_R=\mu_F=\sqrt{m^2_b+p_{\perp,b}^2}$, by a factor of 
$2$ changes the expected cross section by $\pm 10 \, \%$. 
This uncertainty is added in quadrature to the uncertainty due to the 
fragmentation. 
The theoretically expected cross section based on a NLO QCD
 calculation is therefore taken to be $\theory$ to be compared to the 
measured value of $\result$. 
The expectation from the FMNR calculation in LO 
for \xsbb is $0.069\pm0.008$ \nb.
The corresponding LO QCD 
expectation from the {\sc Aroma} generator, with which this measurement 
has been compared previously \cite{vancouver}, is $0.038$ \nb.

\begin{figure}[htb]
  \begin{center}\vspace*{-4mm}
          \mbox{\epsfig{file=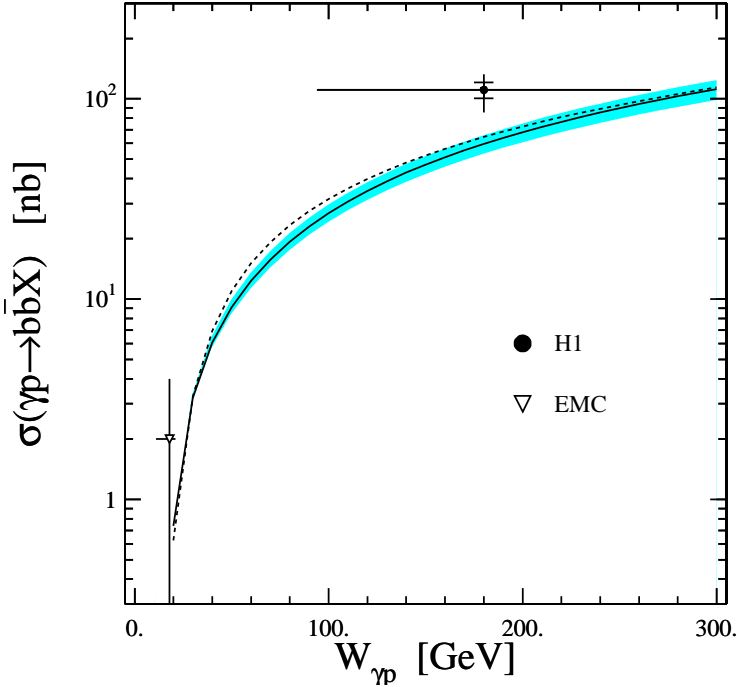,width=0.65\linewidth,clip=
}}\vspace*{-4mm}
\caption{ \small The  total photoproduction
cross section, $\sigma(\gamma p\rightarrow b\bar{b} X)$. The horizontal
error bar represents the range of the measurement. The solid curve shows
  the expectation of the FMNR NLO QCD calculation (full line) with
  $m_b=4.75 \gev$ and the MRSG and GRV-HO structure functions for the
  proton and the photon, respectively. The shaded area corresponds to
  the uncertainty if the factorisation scale changes by a factor of 2.
  A change of the renormalisation scale by a factor of 2 leads to a
  similar result. The dashed line represents the prediction of the
  FMNR NLO QCD calculation if the MRST \cite{mrst} structure function
  for the proton is used. }
    \label{fig:totgp}
  \end{center}
\end{figure}
In the Weizs\"acker-Williams approximation (WWA)\cite{wwa} the
electroproduction cross section, $\sigma_{ep}$, is expressed as a
convolution of the flux of photons emitted by the electron,
$f_{e/\gamma}$, with the photoproduction cross section :
 $$\sigma_{ep}=\sigma(ep\rightarrow e b\bar{b}X)=\int {\mathrm d}y
 f_{e/\gamma}\sigma(\gamma p\rightarrow b\bar{b} X).$$
 In order to
 derive the total photoproduction cross section the measured
 $\sigma_{ep}$ in the visible kinematic range has to be extrapolated
 to the full phase space. 
The fraction of the phase space covered by the defined visible range is 
found to be  12.5\%, from the FMNR NLO calculation with the hadronization
and semileptonic decay, as described above. 
Correcting for the semileptonic branching fractions for a muon 
originating from a $b$ quark \cite{pdg} and extrapolating to the full phase space, the
 total electroproduction cross section for $Q^2<1\gev^2$ is
 $\sigma(ep\rightarrow e b\bar{b}X)=\totep$.
 Using the WWA the  total photoproduction cross section is $\sigma(\gamma
 p\rightarrow b\bar{b} X)=\totgp$  averaged over the range
$94 < W_{\gamma   p}<266 \gev$ with a mean value of  $\langle W_{\gamma
   p}\rangle\sim 180 \gev$.  Here the first and second errors are the
 experimental statistical and systematic uncertainties, respectively.
 The third is the systematic error due to the extrapolation 
  and the branching fraction uncertainties for  
 the semileptonic $b$ quark decays \cite{pdg}.  The expectation of the
 FMNR NLO calculation is 63 \nb. This measurement together with that from EMC 
and the expectation of the FMNR NLO QCD calculation are shown in  figure \ref{fig:totgp}.

\subsection*{Conclusion}
The open $b$ production cross section has been measured for the first
time at HERA using semi-muonic decays of the $b$ quarks. The visible
cross section \xsbb, in the range $Q^2 < 1\gev^2$, $0.1 < y < 0.8$,
$\pt^\mu~>~2.0~\gev$ and $35\grad < \theta^\mu < 130\grad$, is found
to be $\result$,  compared to the expectation
 $0.104~\pm~0.017$~nb from a NLO QCD calculation.  

From this
measurement the total cross section for electroproduction with $Q^2~<~1\gev^2$
 and photoproduction, extrapolated to the full phase space, is
calculated to be $\sigma(ep\rightarrow e b\bar{b}X)=\rtotep$ and
$\sigma(\gamma p\rightarrow b\bar{b}X)=\rtotgp$ at an average
$\langle W_{\gamma p}\rangle\sim 180 \gev$, respectively.  The measured cross
sections are higher than the expectation based on a NLO QCD
calculation.

% ==========================================================================
\subsection*{Acknowledgements}
% =========================
 
We are  grateful to the HERA machine group whose 
outstanding efforts have made and continue to make this
experiment possible. We thank the engineers and technicians 
for their work in constructing and now maintaining the H1
detector, the funding agencies for financial support, the
DESY technical staff for continual assistance, and the
DESY directorate for the hospitality extended to the non-DESY 
members of the collaboration.
We would like to thank S. Frixione for  enlightening discussions 
on the NLO QCD calculation program.

\end{document}